\documentclass[aps,prb,twocolumn,superscriptaddress,citeautoscript]{revtex4-1}
 \usepackage[english]{babel}
 \usepackage{amsmath,bm}
 \usepackage{lipsum}
 \usepackage{amsfonts}
 \usepackage{graphicx}
 \usepackage{setspace} %leaves captions single space in draft mode
 \usepackage{graphicx}
 \usepackage{epstopdf}
 \usepackage{dcolumn}
 \usepackage{amsmath}
 \usepackage{epsfig}
 \usepackage{indentfirst}
 \usepackage{psfrag}
 \usepackage{subfigure}
 \usepackage{amssymb}
 \usepackage{color}
 \usepackage{xcolor}
 \usepackage{siunitx}
 \usepackage{graphicx}% Include figure files
 \usepackage{dcolumn}% Align table columns on decimal point
 \usepackage{bm}% bold math
 \usepackage{natbib}
\usepackage{physics}
\usepackage{dcolumn}% Align table columns on decimal point
\usepackage{bm}% bold mathhttps://www.overleaf.com/project/5af95d0f3775594d14d4a052
\usepackage{natbib}
\usepackage[backref=none,bookmarksnumbered=true,bookmarks=true,bookmarksopen=true,colorlinks=true,
citecolor=blue,linkcolor=blue,anchorcolor=green,urlcolor=blue,unicode=false]{hyperref}

\usepackage{ulem}[normalem] %Whatch out, places underlining in Journal references. Use \normalem before references to deactivate this 

\renewcommand{\vec}[1]{\boldsymbol{#1}}

\makeatletter

\newcommand\colorsout[1]{\bgroup \markoverwith{\textcolor{#1}{\rule[0.5ex]{2pt}{0.4pt}}}\ULon}

\makeatother
\begin{document}

%\title{Theory of the Yu-Shiba-Rusinov excitations in normal metals proximitized by a superconductor}
\title{Theory of a Single Magnetic Impurity on a  Thin Metal Film \\
in Proximity to a Superconductor}
%\title{Single Magnetic Impurity on a Metal Thin Film \\
%in Proximity to a Superconductor}
\author{Jon Ortuzar}
\affiliation{CIC nanoGUNE-BRTA, 20018 Donostia-San Sebasti\'an, Spain}

\author{Jose Ignacio Pascual}
  \affiliation{CIC nanoGUNE-BRTA, 20018 Donostia-San Sebasti\'an, Spain}
\affiliation{Ikerbasque, Basque Foundation for Science, 48013 Bilbao, Spain}

\author{F. Sebastian Bergeret}
 \affiliation{Centro de F\'isica de Materiales (CFM-MPC) Centro Mixto CSIC-UPV/EHU, E-20018 Donostia-San Sebasti\'an,  Spain}
\affiliation{Donostia International Physics Center (DIPC), 20018 Donostia-San Sebastian, Spain}

\author{Miguel A. Cazalilla}
  \affiliation{Donostia International Physics Center (DIPC), 20018 Donostia-San Sebastian, Spain}
\affiliation{Ikerbasque, Basque Foundation for Science, 48013 Bilbao, Spain}

\begin{abstract}
We argue that the formation of Yu-Shiba-Rusinov  excitations in proximitized
thin films is largely mediated by a type of Andreev-bound state named after de Gennes and Saint-James. This is shown by studying an experimentally motivated model and computing the overlap of the wave functions of these two subgap states. We find the overlap stays close to unity even as the system moves away from weak coupling across the parity-changing quantum phase transition. Based on this observation, we introduce a single-site model of the bound state coupled to a quantum spin. The adequacy of this description is assessed  by reintroducing the coupling to the continuum as a weak perturbation and studying its scaling flow using 
Anderson's poor man's scaling. 
\end{abstract}
\date{\today}
\maketitle

\section{Introduction}

The presence of impurities on superconductors results in subgap bound states known as Yu-Shiba-Rusinov (YSR) states~\cite{yu,shiba,rusinov} (see e.g. Ref.~\onlinecite{Balatsky_RevModPhys.78.373} for a review). These  excitations can be probed using scanning tunneling spectroscopy (STS) and appear as narrow resonances in tunneling spectra~\cite{Shuai_YSR,nacho_review}. YSR states were originally discovered as solutions to the scattering problem of a magnetic impurity in bulk superconductors, by treating the magnetic exchange with the impurity as a classical Zeeman field that couples to the local  spin density of quasi-particles~\cite{yu,shiba,rusinov,Balatsky_RevModPhys.78.373}. However, this approximation does not take into account  the quantum nature of the impurity spin, which can give rise to many-particle effects such as the Kondo effect~\cite{Kondo} and it is determinant when e.g. describing the spin carried by the YSR excitations~\cite{skurativska2023robust}. 

A fully quantum-mechanical treatment of this problem aimed
at providing a comprehensive description of experiments~\cite{carmen_YSR},  
often requires the use of sophisticated but numerically costly methods such as the numerical renormalization  
group (NRG) \cite{NRG_YSR,zitko} or continuous time MonteCarlo~\cite{Odobesko_PhysRevB.102.174504}.
\begin{figure}[b]
    \centering
    \includegraphics[width=0.45\textwidth]{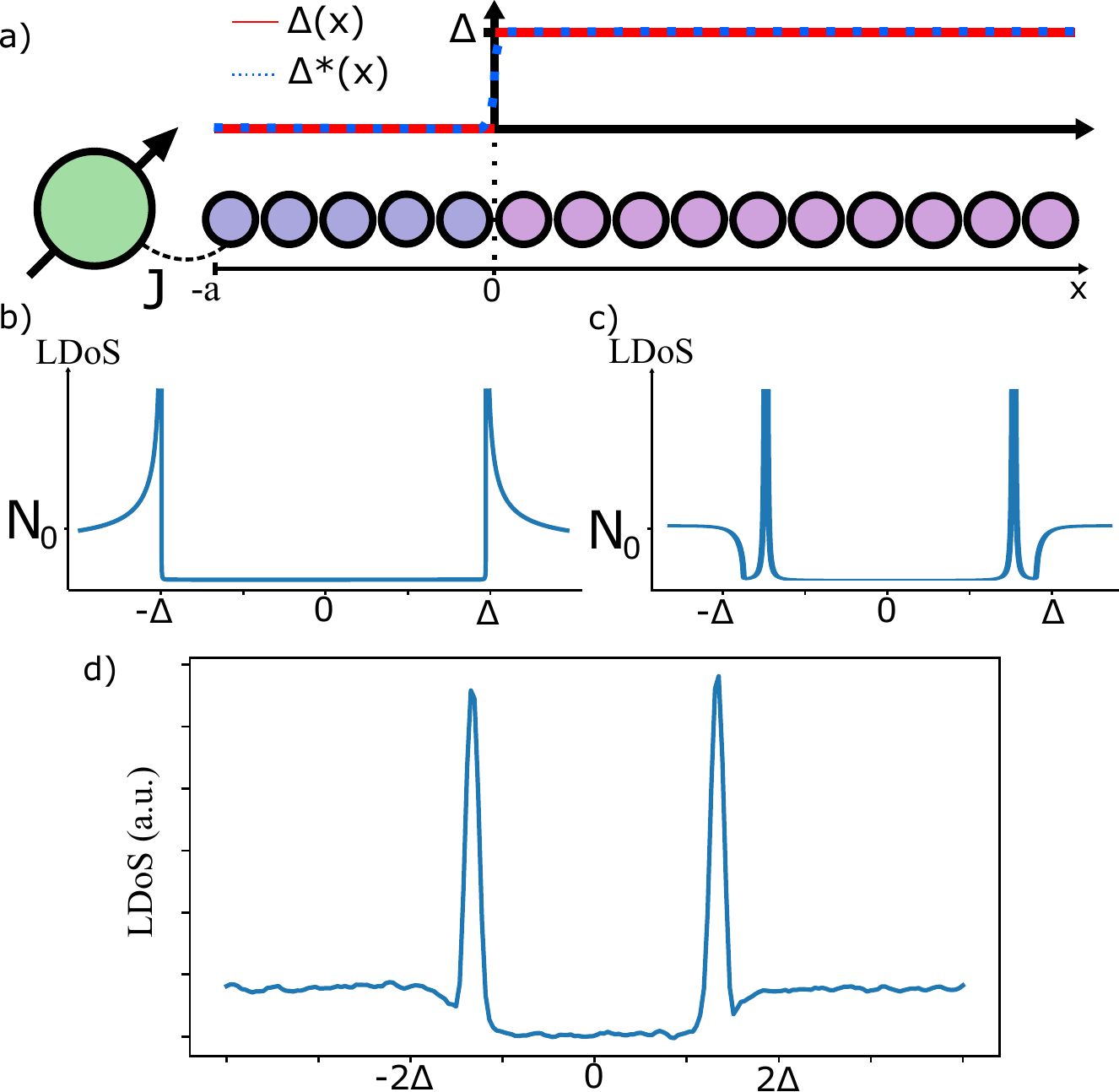}
    \caption{Panel (a) shows a sketch of the studied system: a magnetic impurity interacting via exchange  with a proximitized thin metal film. Panels (b) and (c) are the local density of states (LDoS) in the bulk of the superconductor and in the thin proximitized metallic film. Panel (d) is a convoluted \cite{nacho_review} spectroscopic measurement using STM of a thin (about 4 mono-layers) proximitized Au film on V($100$) }
    \label{fig:1}
\end{figure}
In recent years,  single-site models \cite{affleck_andreev_2000,vecino2003josephson,OppenModel} have emerged as  a computationally affordable  approach to treat some of the quantum many-particle aspects of the YSR problem~\cite{OppenModel}. These models have already been successfully used to  explain some spectral features observed in recent experiments~\cite{PhysRevLett.130.136004,eva_shiba_chain,zeeman_YSR}.
Moreover,  it has been also  applied to explain the complex many-body physics of a magnetic molecule on a clean gold film proximitized by a superconducting substrate\cite{PhysRevLett.130.136004}. 

Proximitized systems have  been studied mainly in the diffusive limit using the Usadel formalism \cite{PhysRevLett.25.507}. This approach predicts the decay of the proximity effect  as well as spectroscopic features such as the closing of the gap  and the formation of a 
minigap~\cite{SNS_DoS,proximity_squid}. Experiments with diffussive systems ~\cite{dirty_proximity,proximity_exp_moodera,roditchev_proximity,proximity_squid} have clearly confirmed
those predictions. However,  the systems studied in this context are mesoscopic in size and the experimental probes that have been employed cannot resolve the behavior of a single magnetic impurity.

  On the other hand, thanks to currently available growth techniques, it is possible to grow clean metallic overlayers with thicknesses of few atomic layers on top of superconductors~\cite{PhysRevLett.130.136004,katerina,lucas,jung}. 
These novel hybrid systems open the door to otherwise impossible on-surface synthesis, and may allow one day  the study of self-organized spin chains~\cite{Zhao_spin_chain,Mishra_spin_chain,jeremy} as well as other,  more complex, molecular structures~\cite{triangulene_dimmers} on superconductors. Such systems are clearly not in the diffusive limit and have to be described within the ballistic limit.
In this case, subgap bound states appear in the normal region and extend into the superconductor over distances of the order of the coherence length. The existece of such states has been known for some time, since the work of de Gennes and Saint James~\cite{SJdG,kulik}.

The aim of this work is to extend the application of the single-site model~\cite{OppenModel} to describe the complexity of a single magnetic impurities interacting with a thin metallic film in proximity to a superconductor. This is a problem of much interest to a number of recent experiments~\cite{PhysRevLett.130.136004,eva_YSR_graphene,proximity_YSR_exp}. Below, 
we first study the   system   treating the magnetic impurity as a classical spin in the ballistic limit where there is a single de Gennes-Saint James (dGSJ) bound state in the gap. We find that a large overlap exists between  the wavefunctions of the dGSJ and YSR states.
Motivated by this result, we propose that the single-site model is a relevant simplified model for complex
system consisting of the magnetic impurity on the proximitized thin film.  The model can be solved exactly and also provides a computationally  cheap way to treat the many-particle effects associated with the quantum spin of the impurity.  The adequacy of the single-site model for the system of interest here is  assessed by means of a "poor man's" scaling analysis. To this end, we introduce a Hamiltonian consisting of a single-site model perturbed by an impurity-mediated coupling to the continuum of other excitations. Under certain conditions we find that, as the high-energy continuum states are integrated out, the impurity remains most strongly coupled to the single site describing the dGSJ state.

The structure of the article is the following: In Sec.~\ref{sec:SaM} we describe the system and  the approximations used.  Section \ref{sec:classical} is divided in two subsections with the first focusing on  the YSR states resulting from the interaction of the magnetic impurity with the proximitized film. This study is undertaken assuming the spin of the magnetic impurity can be treated classically.   In the second subsection, we describe the calculation of the wave function overlap between the YSR and dGSJ states.  In Sec.~\ref{sec:singlesite}, we introduce the single-site model for the magnetic
impurity on a proximitized film.  Finally, in section \ref{sec:PoorMan}, we argue that the single-site model provides an accurate description of this system using Poor man's scaling\cite{Anderson_poorman}.
The most technical details of the calculations have been relegated to the Appendices.

\section{ System and Model}\label{sec:SaM}

Fig. \ref{fig:1} shows a schematic picture of the system studied in this work, which is motivated by experiments reported in Ref.~\onlinecite{PhysRevLett.130.136004} and Refs.~\onlinecite{katerina,lucas,jung}. The system consists of a magnetic impurity on top of a thin normal metal film (N) in proximity to a superconductor (S). The superconductor occupies the half-space $x>0$, while the N film corresponds to $-a<x<0$. The system is translationally invariant in the $(y,z)$-plane, so it is convenient to describe the electron  wave function as  $\psi(x,\mathbf{k_{\parallel}})$, where $\mathbf{k_{\parallel}}$ is the component of the momentum vector parallel to the S/N interface at $x=0$. We assume a perfect S/N interface with no Fermi surface mismatch or potential barrier, such that the N region acts as a cavity for electrons with energy $E<\Delta$: they undergo Andreev retro-reflections at the S/N interface and normal specular reflections at the interface with vacuum. According to the Bohr-Sommerfeld quantization rule, the phase accumulated along a closed classical trajectory must be a multiple of $2\pi$. In the N/S system under consideration, a closed trajectory consists of two Andreev retro-reflections at the S/N interface and two normal reflections at $x=-a$. Thus,
\begin{equation}
\label{eq:dGsJ_0}
    \frac{2aE}{\hbar v_F\cos\varphi}-\cos^{-1}\left(\frac{E}{\Delta}\right)=n\pi
\end{equation}
where $\cos^{-1}(E/\Delta)$ is the phase shift associated to each AR, and $\cos\varphi={k_\parallel/k_F}$. 
Eq.~\eqref{eq:dGsJ_0},  determines the subgap bound states, also known as De Gennes-Saint James (dGSJ) states \cite{SJdG}.
It is valid for clean N-layers with a mean free path larger than the thickness $a$, and it describes a continuum of subgap states~\cite{SJdG,deGennes_Boundary}. 

 For the STM experiments of interest to us here, assuming specular tunneling\cite{McMillan},   the decay of the wavefunction of these excitations in vacuum is determined by the metal work-function. This energy scale is of the order of one electron-volt and therefore much larger than the superconducting gap. Therefore, in vacuum, the tail of the dGSJ wave function is essentially indistinguishable from that of an electron at the Fermi level  in the normal state, and excitations with finite $\vec{k}_{\parallel}$  penetrate less into the vacuum. As a result, when probed with a STM in the tunneling regime,   excitations with large $|\vec{k}_{\parallel}|$ are filtered out~\cite{kieselmann1987self,arnold,McMillan} and dGSJ states are observed as  narrow subgap resonances made of dGSJ quasi-particles with $\vec{k}_{\parallel}\approx \vec{0}$~\cite{PhysRevLett.130.136004}. Moreover, a small amount of disorder will randomize trajectories 
with $\cos\varphi<a/l$, where $l$ is the mean free path, suppressing the coherence of such trajectories. 
A magnetic impurity  on top of the proximitized film has compact and anisotropic orbitals that typically couple to several scattering channels from the substrate. However, since the dGSJ quasi-particles with $\vec{k}_{\parallel} \simeq \vec{0}$ penetrate  farther into  the vacuum, they are also expected to contribute substantially to the most strongly coupled scattering channel. 
Thus, one can effectively approximate the tunneling problem  using a one-dimensional model which  neglects the motion parallel to the surface: 
\begin{equation}\label{H}
     \mathcal{H}= \mathcal{H}_0 +\mathcal{H}_J\; , 
\end{equation}
where
\begin{equation}\label{H0}
\begin{split}
    \mathcal{H}_0=& \sum_{\sigma} \int_{-a}^{\infty} dx\, \psi^{\dagger}_{\sigma}(x)  \left[ -\frac{\hbar^2}{2m^*}\partial^2_{x}-E_F \right] \psi_{\sigma}(x^{\prime})\\
    &+\int_{0}^{\infty} dx \, \Delta\: \psi^{\dagger}_{\uparrow}(x)\psi_{\downarrow}(x)+ \mathrm{h.c.}\; , 
\end{split}
\end{equation}
and
\begin{equation}\label{HJ}
    \mathcal{H}_J=\sum_{\sigma\sigma'}J \psi^{\dagger}_{0\sigma} \vec{S}\cdot\vec{s}_{\sigma\sigma'}\psi_{0\sigma'}\; .
\end{equation}
Here, $\psi_{\sigma}(x)$ ($\psi^{\dag}_{\sigma}(x)$) represents the annihilation (creation) operator for an electron with spin $\sigma = \uparrow,\downarrow$ in the metal-superconductor substrate. $H_0$ describes a proximitized thin film of thickness $a > 0$. The first term contains the kinetic energy and chemical potential $E_F$, and the second term is the s-wave pairing potential. The pairing potential is not self-consistently calculated. Corrections due to self-consistency  result in a spatially non-uniform pairing potential $\Delta(x)$, but they have only a small effect on the spectral properties of the dGSJ  states~\cite{arnold,quas_arnold}. The magnetic exchange with the impurity is described by $H_J$, with $\vec{s}$ denoting the electron-spin Pauli matrices and $\vec{S}$ denoting the impurity spin operator. The operators $\psi_{0\sigma}$ ($\psi^{\dag}_{0\sigma}$) annihilate (create) electrons at the position of the impurity. For the one-dimensional model introduced above, $\psi_{0\sigma} = \psi_{\sigma}(x=-a)$. In the following section, we analyze this model using the approach of Yu, Shiba, and Rusinov (YSR)~\cite{yu,shiba,rusinov}, where the impurity spin $\vec{S}$ is treated as a classical vector. 

\section{YSR  in Proximitized Thin Films}\label{sec:classical}

In the previous section, we have derived the equation that determines the  spectrum of subgap states (cf Eq. \ref{eq:dGsJ_0}) using the Bohr-Sommerfeld semiclassical approximation. As explained above, we will focus on the one-dimensional case, which corresponds to $\cos\varphi=1$ in Eq.~\eqref{eq:dGsJ_0}. To deal the coupling to the magnetic impurity, we  solve the model described by Eqs.~\eqref{H}-\eqref{HJ}. To this end, we use Green's functions (GFs) and follow the approach outlined in Ref.~\onlinecite{arnold}. The technical details of the calculation are described in Appendix~\ref{app:A}. From the knowledge of the retarded GFs, $G(\omega+i\eta,x,x)$, the local density of states (LDoS) $\rho(\omega,x)$ of the system is obtained by using $\rho(\omega,x)=-\tfrac{1}{\pi}\Im\: G(\omega+i\eta,x,x)$. 

Fig. \ref{fig:2} (a) shows the LDoS on the surface as a function of film thickness. As we increase the thickness, new dGSJ states enter the gap. The GF also has poles with a finite imaginary part outside the superconducting gap that correspond to states in the continuum (i.e. above the superconductor gap), and give rise to McMillan-Rowell-Tomasch oscillations~\cite{rowell1966electron,tomasch1966geometrical}. From here on, we focus our discussion on thin films with a single subgap bound state. In the following subsection, we tackle the coupling  to the magnetic impurity.

\subsection{YSR states}

The  GF for the S/N system provides the starting point for calculating the spectral properties of the YSR excitations. The properties of the latter  can be obtained by solving the following integral equation:
\begin{equation}\label{GYSR}
\begin{split}
    G_{YSR}(x,x')=&G(x,x')+G(x,-a)V\times\\
    &(1-VG(-a,-a))^{-1}G(-a,x')\; .
\end{split}
\end{equation}
Here, $G(x,x')$ is the GF obtained in the previous subsection.  The scattering potential for a spin-$S$ impurity in the Nambu notation is $V=JS \sigma_z \tau_0$, assuming that the impurity (classical) spin points along the $z$-axis.

\begin{figure}
    \centering
    \includegraphics[width=0.45\textwidth]{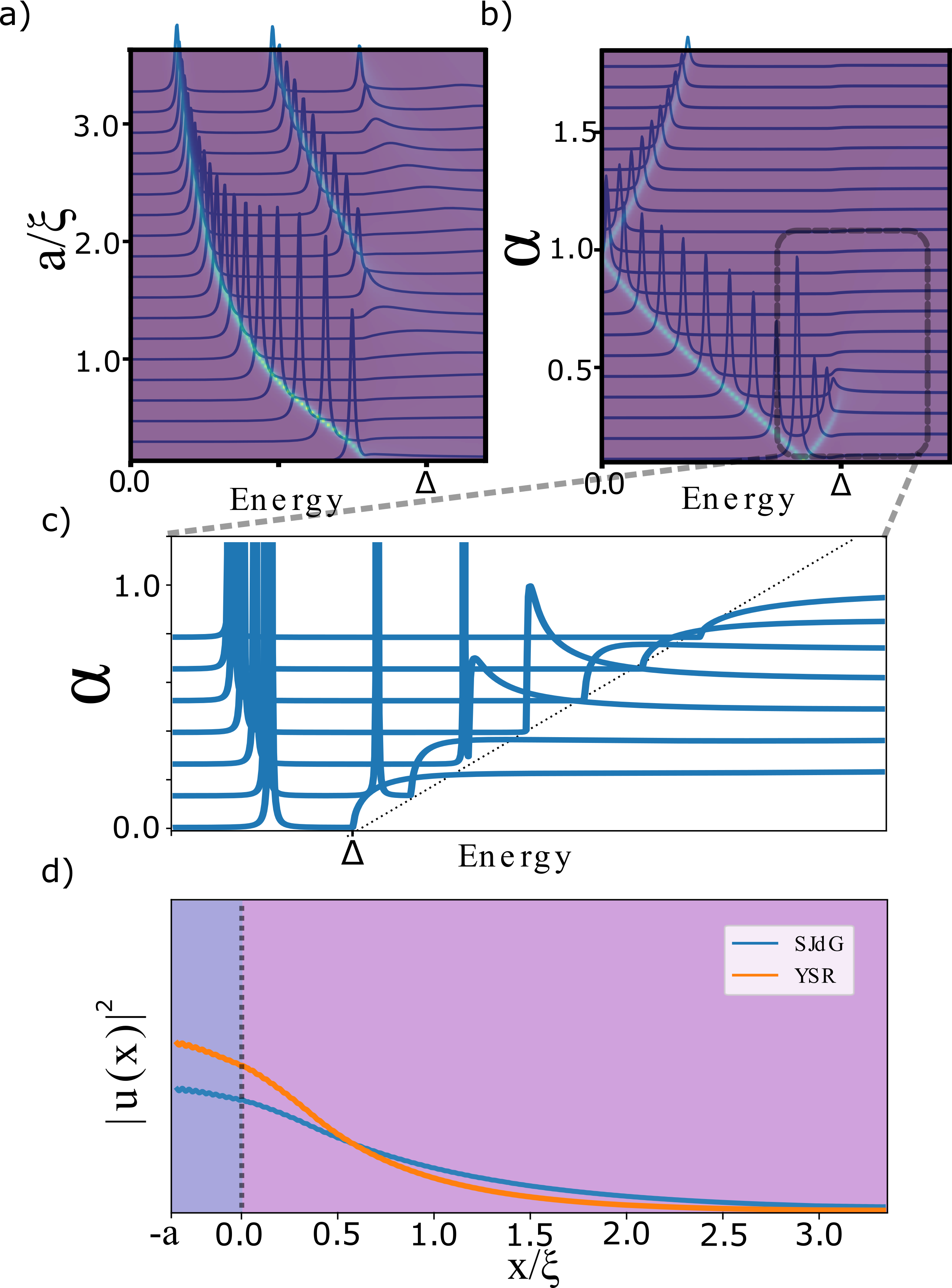}
    \caption{(a) Evolution of the dGSJ states as a function of the thickness of the metallic layer. (b) Evolution of the YSR state for a fixed metallic layer thickness as a function of the exchange coupling. (c) Zoom-in of the evolution of the YSR state. (d) Wavefunction of the SJdG and YSR states averaged over distances $\gg k^{-1}_{F}$.}
    \label{fig:2}
\end{figure}

Fig \ref{fig:2} (b) shows the evolution of the YSR state as a function of the exchange coupling $\alpha=\nu_0\pi J S$, with $\nu_0$ being the normal metal DoS  defined so that the quantum phase transition (QPT), where the energy of the YSR state crosses the center of the gap, happens for $\alpha=1$\cite{yu,shiba,rusinov}. Note that the exchange coupling splits dGSJ  state into two states (spin up and down), one of which shifts to higher energy while the other shifts to lower energy, see Fig.~\ref{fig:2}(c). As $J$ increases beyond a certain value, the higher energy state disappears into the continuum. From this point on, the energy of the remaining subgap state behaves similarly to a YSR in a bulk superconductor \cite{yu,shiba,rusinov}. For thicker films, with more than one dGSJ state, the behavior is similar: each bound state splits in two, shifting in opposite directions depending on their spin projection, with more excited states eventually merging in the continuum and disappearing.

The transmutation of the dGSJ into the  YSR state can be regarded as a consequence of a spectral reorganization taking place around $|\omega| = \Delta$ caused by AR (see Fig.~\ref{fig:1}c). Using an analogy to semiconductor physics, YSR states appear in a superconductor because the coherence "peak" behavior $\sim (\omega^2-\Delta^2)^{-1/2}$ (cf. Fig.\ref{fig:1} b) resembles a van Hove singularity at the bottom (top) of the conduction (valence) band of a one-dimensional insulator. Bound states appear due to the infinitesimal attraction provided by the magnetic impurity Dirac-delta potential. However, in a proximitized film, AR reorganizes the spectral weight by removing the van Hove-like singularity while shifting most of its spectral weight to the dGSJ state (cf. Fig.\ref{fig:1} c). Together with the localization of the dGSJ states at the surface, this enables the transmutation of one of the dGSJ states per spin into a YSR. Thus, a large overlap of the wavefunctions of YSR and dGSJ states is expected, as explicitly demonstrated in the following subsection.

\subsection{Overlap between SJdG and YSR States}

In this section,  we compute  the  overlap of the YSR  and the dGSJ states as a function of the exchange coupling $J$.
This can be achieved by using the GF obtained from the scattering solution of the problem with and without magnetic impurity. The square of the overlap is computed from the following integral involving the residue  of the two GFs:
\begin{align}\label{overlap}
|\Theta|^2 &=\int dx \left[ u_{dGSJ}(x)u^*_{YSR}(x) + v_{dGSJ}(x)v^*_{YSR}(x) \right]\notag \\
    &= \int dx dx' \, \mathrm{Tr}\left\{  \Res\: G(x,x')) \Res\: G_{YSR}(x',x) \right\}.
\end{align}
Here $\Res G_{YRS}$ ($\Res G$) is the residue of the Nambu GF matrix at the YSR (dGSJ) pole with spin up. 

\begin{figure*}
    \centering
    \includegraphics[width=\textwidth]{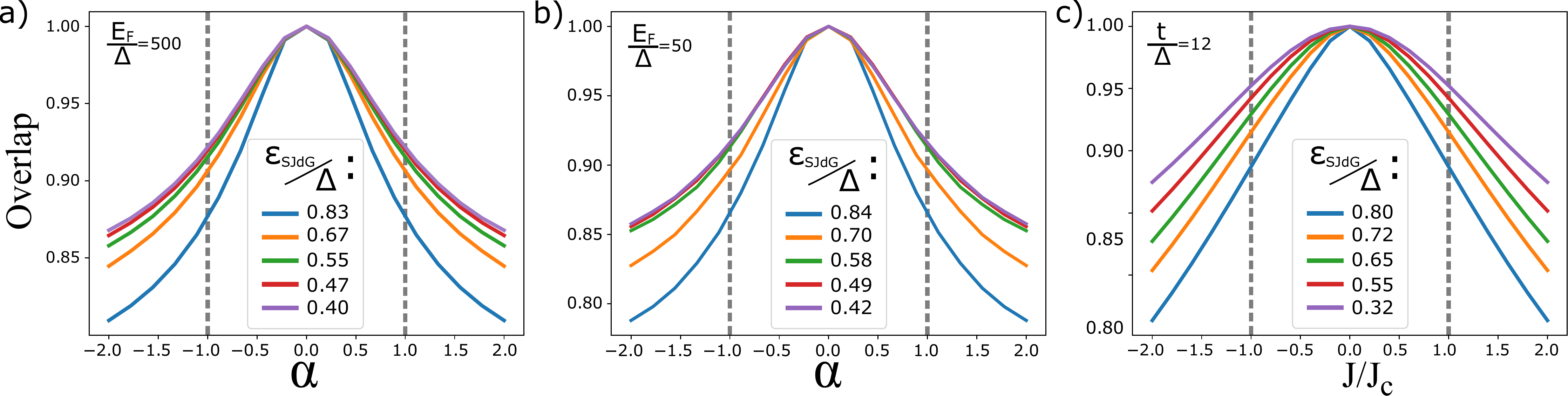}
    \caption{Overlap between the YSR wavefunction and the SJdG wavefunction. Panels (a) and (b) show the calculation done from the continuous model for different SJdG bound state energies and values of $E_F/\Delta$. Panel (c) shows the same calculation done with a tight-binding model.}
    \label{fig:3}
\end{figure*}

In Fig.~\ref{fig:3} (a) and (b)  we show the behavior of the overlap $\Theta$  as a function of exchange coupling $J$ for different values of film thickness (which determines the dGSJ state energy).
To check our results beyond the leading order in $\Delta/E_F$, we also compute the overlap by solving the Bogoliubov-de Gennes equations for a one-dimensional tight-biding chain containing up to $2000$ sites. The results are shown in Fig.~\ref{fig:3}(c) as a function of $J$ normalized to the critical value $J_c$ where the system undergoes the parity-changing quantum phase transition~\cite{yu,shiba,rusinov,Balatsky_RevModPhys.78.373}. 
The overlap between the YSR and dGSJ states decreases as the exchange coupling increases, but it remains close to unity even across the quantum phase transition. It is worth noting that the energy of the YSR excitation shifts away from that of the dGSJ state as the exchange coupling is increased. The significant overlap between the two states suggests that the YSR state primarily descends from the dGSJ state, with  a minor contribution from the continuum states of the proximitized film. Therefore, in a first approximation, the coupling with the magnetic impurity can be described  by replacing the proximitized film with a single level representing the dGSJ state.

\section{Single-site Model}\label{sec:singlesite}

Motivated by  the results of the previous section, we introduce a simplified model that replaces the entire proximitized film with a single site representing the dGSJ state. As we show below,  this model is useful for analyzing the coupling between the dGSJ state and a quantum spin. The Hamiltonian of the single site is given by:
\begin{equation}
H_0 = \sum_{\sigma} E_s \left( \gamma^{\dag}_{\sigma} \gamma_{\sigma}  - \frac{1}{2}\right),
\end{equation}
where $\gamma_{\sigma}$ ($\gamma^{\dag}_{\sigma}$) are the annihilation (creation) operators for a dGSJ quasi-particle with spin $\sigma = \uparrow, \downarrow$, and $E_s$ is the  eigenvalue of the BdG Hamiltonian (in the absence of magnetic impurity). As explained in Appendix~\ref{app:B}, this Hamiltonian can be recast in terms of electron operators $d_{\sigma},d^{\dag}_{\sigma}$ as follows:
\begin{equation}
H_0 = U \sum_{\sigma} n_{\sigma} + \left[ \Delta_s d_{\downarrow} d_{\uparrow} + \mathrm{h.c.} \right],
\label{eq:H0Sd}
\end{equation}
where $n_{\sigma} = d^{\dag}_{\sigma}d_{\sigma}$;  $U$ and $\Delta_s$  are effective
scattering and pairing potentials, respectively. In terms of $U$
and $\Delta_s$, $E_s = \sqrt{U^2 + \Delta^2_s}$. Without loss of generality, below we discuss the particle-hole symmetric case  where $U = 0$ and therefore $E_s = \Delta_s$

Next, we introduce the coupling to the impurity. To make contact with the classical description employed in the previous section, we first discuss the Ising limit of the exchange coupling, i.e.
\begin{equation}
H^{\mathrm{Ising}}_J = J^{\parallel}_{dd}   S^z \left(n_{\uparrow}-n_{\downarrow}\right),\label{eq:ising}
\end{equation}
where $J^{\parallel}_{dd} > 0$ is the exchange coupling with the dGSJ  quasi-particle.  This model reproduces the most salient features of the YSR states described above. To begin with, note that, besides
the fermion parity  $P = \prod_{\sigma}(-1)^{n_{\sigma}} = \pm 1$, 
the impurity spin operator $S^z$ is  also conserved in this limit, i.e. $\left[ S^z, H_0 + H^{\mathrm{Ising}}_J \right] = 0$. Thus, the ground state is  doubly degenerate corresponding to the
two possible orientations of the classical vector $\vec{S} = \pm S \vec{\hat{z}}$: For $J_{dd} < J_c = 2 \Delta_s$  the ground state is one  of the two following  states $\{ |BCS\rangle \otimes |\pm \tfrac{1}{2}\rangle \}$ with $P = +1$  and $ \gamma_{\sigma}|BCS\rangle = 0$. For $J_{dd} > J_c$, the ground state is one  in $\{|\uparrow\rangle\otimes|-\tfrac{1}{2}\rangle, |\downarrow\rangle\otimes|+\tfrac{1}{2}\rangle \}$ with $P=-1$ and $|\sigma\rangle = \gamma^{\dag}_{\sigma}|BCS\rangle $. The YSR excitation is a transition between these two  ground states of opposite parity with excitation energy~\cite{OppenModel,PhysRevLett.130.136004} $|\Delta_{s} - J_{dd}/2|$. In addition, the odd parity sector of the Hilbert space also contains the following two states: $\{ | \uparrow\rangle\otimes |+\frac{1}{2}\rangle,  | \downarrow \rangle\otimes |-\frac{1}{2}\rangle \}$ with  excitation energy equal to $\Delta_{s} + J_{dd}/2$. For small $J_{dd}$, a transition from the  ground state with $P=+1$ to  these states corresponds to the second subgap peak in the LDoS of the classical approach that shifts up in energy with increasing exchange and eventually disappears into the continuum, see Fig \ref{fig:2} (c).

  Next, we generalize Eq.~\eqref{eq:ising} by adding the spin-flip term, which allows 
the impurity spin  to fluctuate:
\begin{equation}\label{singlesite} 
H^{d}_{J}= J^{\parallel}_{dd}   S^z \left(n_{\uparrow}-n_{\downarrow}\right) +  J^{\perp}_{dd}(S^{+} d^{\dag}_{\downarrow}d_{\uparrow} + \mathrm{h.c.})
\end{equation}
 As argued in  Refs.~\onlinecite{OppenModel,vecino2003josephson}, the single-site model provides an economical and fully quantum-mechanical description of YSR spectra in superconductors which compares well with the results obtained using  sophisticated but computationally expensive methods like the Numerical Renormalization Group (NRG)~\cite{zitko}. The accuracy of this description  in the present system 
 will be addressed in the following section.
 
 The spin-flip term, $\propto J^{\perp}_{dd} > 0$, has important consequences for the spectrum of the model.  In the weak coupling  limit, i.e. for $J^{\parallel}_{dd} + 2J^{\perp}_{dd} 
 < 2 \Delta_s$, (assuming an unbiased  preparation of the system) 
the ground state is described by the following density matrix:
 \begin{widetext}
 \begin{equation}
 \rho_{GS} = \frac{1}{2} \left[ |BCS, +\tfrac{1}{2}\rangle \langle   \langle +\tfrac{1}{2},BCS|  + |BCS, -\tfrac{1}{2}\rangle  
 \langle - \tfrac{1}{2},BCS| \right],
 \end{equation}
 \end{widetext}
On the other hand, in the strong coupling limit  where  $J^{\parallel}_{dd} + 2J^{\perp}_{dd}   > 2 \Delta_s$, the ground state is a singlet:  
 \begin{equation}
 |GS \rangle  = \frac{1}{\sqrt{2}} \left(|\uparrow\rangle \otimes |+\tfrac{1}{2}\rangle + |\downarrow\rangle \otimes |-\tfrac{1}{2} \rangle \right).  
 \end{equation}
that is, a pure state resulting from the quantum superposition of the two ground
states of the Ising limit of the model.
In weak and strong-coupling regimes,  unlike the conventional classical approach of YSR~\cite{yu,shiba,rusinov},  the quantum model predicts that YSR excitations carry 
no spin polarization~\cite{skurativska2023robust}.

 Finally, since in the original model (cf. Eq.~\ref{HJ}) the energy of the YSR 
 does not grow without bound as the exchange with the magnetic impurity  $J$ becomes arbitrarily large,
 the couplings $J^{\perp}_{dd},J^{\parallel}_{dd}$ cannot be much larger than $\Delta_s$ 
 in the single-site model. Note that, for large   $J^{\perp}_{dd},J^{\parallel}_{dd}$
 the energy of the YSR grows like $\max\{J^{\perp}_{dd},J^{\parallel}_{dd}\}$. 
 Thus, for the energy of the YSR to remain within the gap, the exchange couplings of the single-site model must saturate to an upper bound so that $\max\{J^{\perp}_{dd},J^{\parallel}_{dd}\}  \lesssim \Delta_s$. Therefore, they   must be regarded as renormalized exchange interactions, which are also
 the result of the spectral reorganization and  localization of excitations with energy $\sim \Delta_s$ 
 caused by Andreev reflection at the S/N interface.

%Comparing results with the ones obtained via numerical renormalization group calculations \cite{zitko}, F. V. Oppen and franke \cite{OppenModel} showed that the physics of the Kondo Hamiltonian for a superconductor can be qualitatively explained within the framework of a single-site model,
%\begin{equation}\label{singlesite} \mathcal{H}=J\sum_{\sigma\sigma'}c^{\dagger}_{\sigma}\vec{S}\cdot\vec{s}_{\sigma\sigma'}c_{\sigma'}+\Delta c^{\dagger}_{\uparrow}c^{\dagger}_{\downarrow}+h.c.\; .
%\end{equation}

%This simplification can be interpreted as saying that most of the spectral weight of the YSR state is coming from a short energy range around the coherence peak of the superconductor. This makes it possible to integrate-out the contribution of the continuum. The aim of this section is to argue that Eq. \eqref{singlesite} can be used to model an impurity on a proximitized superconductor just by exchanging $\Delta$ for $E_{SJdG}=\tilde\Delta$. As shown in Section \ref{sec:classical}, the overlap between the YS state and the Andreev bound state is big. This points out that the YSR spectral weight is coming primarily from the dGSJ state, which leads us to believe that our hypothesis is correct.

\section{Scaling Approach}\label{sec:PoorMan}

In order to investigate the accuracy of the single-site model, we  reintroduce the coupling to the
continuum of excitations as a perturbation. 
Whether this perturbation  changes  the low-energy spectrum substantially or not can be assessed using the poor man's scaling method~\cite{Anderson_poorman}, as we describe in the following. 

In the single-site model, the effective exchange coupling of the impurity and dGSJ quasi-particle is $J_{dd} = J^{\perp}_{dd} = J^{\parallel}_{dd}$, where, for the sake of simplicity, we assume an isotropic coupling.  Our conclusions also apply to the anisotropic case with small modifications.  Through the exchange interaction with the magnetic impurity, the dGSJ quasi-particles can also couple to the continuum of excitations of the proximitized film.  Let us introduce the following modified exchange coupling which, besides the coupling to the dGSJ,  describes an impurity-mediated coupling of the dGSJ-site to the continuum, and  will be treated below as a perturbation:
\begin{equation}
\begin{split}
H^{dc}_{J}&= \sum_{\sigma\sigma^{\prime}}\left( J_{dd}\:  d^{\dag}_{\sigma} \vec{s}_{\sigma\sigma^{\prime}} d_{\sigma} + J_{\Phi\Phi}\: \Phi^{\dag}_{0\sigma} \vec{s}_{\sigma\sigma^{\prime}} \Phi_{0\sigma^{\prime}}  \right) \cdot \vec{S}\\
&\qquad + J_{d\Phi} \sum_{\sigma\sigma^{\prime}}\left( d^{\dag}_{\sigma} \vec{s}_{\sigma \sigma^{\prime}}  \Phi_{0\sigma^{\prime}} 
+  \Phi^{\dag}_{0 \sigma} \vec{s}_{\sigma \sigma^{\prime}}  d_{\sigma^{\prime}}   \right) \cdot \vec{S}. \label{eq:phenolham}
\end{split}
\end{equation}
The operators $\Phi_{0\sigma},\Phi^{\dag}_{0\sigma}$ are the annihilation and creation operators for electrons in the continuum at the position of magnetic impurity. Phenomenologically, we have assumed different couplings for the various processes involving the scattering of the dGSJ  and the continuum excitations by the impurity. These couplings can be calculated from first principles. However, they depend on  microscopic details of the matrix elements of the impurity orbitals and the continuum of both subgap and outer-gap excitations which are difficult to model. For this reason, we  treat their bare values as free parameters in the analysis below. 

 We carry out the poor man's scaling  analysis~\cite{Anderson_poorman} of the model~\eqref{eq:phenolham}  by integrating out the high energy degrees of freedom from the continuum with energies of the order of the bandwidth $D\sim E_F$.  Since these band-edge modes exhibit vanishing superconducting correlations because
their energies are well above the gap, the calculations do not differ
much from those of the standard Kondo scaling of a magnetic impurity~\cite{Anderson_poorman}. Some details are provided in Appendix~\ref{app:C}. In what follows, we focus on the discussion of the solutions to the  scaling equations, which read
\begin{align}
\frac{dg_{\Phi\Phi}}{d\ell} &=  g^2_{\Phi\Phi}, \label{eq:rg1}\\
\frac{dg_{d\Phi}}{d\ell} &=  g_{d\Phi} g_{\Phi\Phi},  \label{eq:rg2}\\
\frac{dg_{dd}}{d\ell} &=  g^2_{d\Phi}.  \label{eq:rg3}
\end{align}
Here $g_{dd} = 2 \nu_0 J_{dd}$,
$g_{d\Phi} = 2\nu_0 J_{d\Phi}$, and $g_{\Phi\Phi} = 2\nu_0 J_{\Phi \Phi}$  are dimensionless couplings, $\nu_0\sim 1/D$ being the mean density of continuum states.  The scaling variable $\ell$  is defined such that the bandwidth is reduced according to 
$D(\ell) = D e^{-\ell}\to 0$ as $\ell \to +\infty$, where $D\sim E_F$.  

As the  bandwidth of the system is reduced, the above scaling equations imply that the renormalization of $g_{dd}$ and $g_{d\Phi}$ is driven by the growth of $g_{\Phi\Phi}$. Indeed, Eq.~\eqref{eq:rg1} for $g_{\Phi\Phi}$ is mathematically identical to the scaling equation  for the exchange coupling of a magnetic impurity in a normal metal (Kondo scaling).   It can be readily solved by the ansatz $g_{\Phi\Phi}(\ell) = (\ell^*-\ell)^{-1}$, where $\ell^* = 1/g_{\Phi\Phi}(0)$. Like  the ordinary Kondo scaling, $\ell^*$ corresponds to the logarithmic scale where  $g_{\Phi\Phi}(\ell)$  diverges and the perturbative renormalization breaks down. This happens when the bandwidth   becomes of the order of a ``Kondo temperature'', $T^{\Phi}_K$, i.e.  for $\ell^* = \log(D/T^{\Phi}_K)$.  Hence, $g_{\Phi\Phi}(\ell^*)\sim 1$ leads to $T^{\Phi}_K = D e^{1/(2\nu_0 J_{\Phi\Phi})}$. 
Note that $T^{\Phi}_K \gg \Delta_s$  would imply that the continuum states at energies  much higher than the superconducting gap are strongly coupled to the magnetic impurity. In this situation,  the single-site description as introduced above breaks down. In the classical approach, such a strong coupling to the continuum should  result in substantial suppression of the overlap between the YSR and dSGJ states.

Indeed,   the wavefunction overlap $\Theta$ (cf. Fig.~\ref{fig:3}) can be used to  obtain a rough estimate the ratios of the bare couplings $ g_{d\Phi}(0)/g_{dd}(0)$, and $g_{\Phi\Phi}(0)/g_{dd}(0)$. To this end, we first notice that $g_{dd} \sim J_{dd}$, $g_{d\Phi}\sim J_{d\Phi}$, and $g_{\Phi\Phi}(0)\sim J_{\Phi\Phi}$
 contain matrix elements with zero, one, and two powers of the continuum orbitals, respectively (recall that the exchange couplings are second order in the  matrix element describing the tunneling between the impurity magnetic orbital and the metallic host states). Let $\gamma = 1-|\Theta|$ measure the degree of admixture of the YSR state with the continuum;  $\gamma$ will be  enhanced by quantum fluctuations relative to the estimates provided by the classical approach (cf. Sec.~\ref{sec:classical}). Nonetheless,  we expect $\gamma$ to remain much smaller than one. Thus,
 $g_{dd}(0)\sim \gamma^0$,  
 $g_{d\Phi}\sim \gamma$ and $g_{\Phi\Phi}\sim \gamma^2$, 
 to leading order in $\gamma$.
Furthermore,  $g_{dd}(0) = 2\nu_0 J_{dd}\sim \Delta_s/D \sim \Delta/D\ll 1$ according to the discussion
at the end of the previous section. 

 Next, we proceed to obtain solutions to the scaling equations using the above estimates
 for the initial conditions of the flow. Concerning the solutions of 
 \eqref{eq:rg2} and \eqref{eq:rg3}, we notice that 
 \eqref{eq:rg2}  is solved by the ansatz $g_{d\Phi}(\ell) = r_{d\Phi}/(\ell^*-\ell)$ with
$r_{d\Phi} = g_{d\Phi}(0)/g_{\Phi\Phi}(0)$. Introducing this result into
Eq.~\eqref{eq:rg1} and  integrating,  we obtain the following renormalized coupling between the 
impurity and the  dSGJ:
\begin{equation}
g_{dd}(\ell) = g_{dd}(0) +  \frac{g^2_{d\Phi}(0)}{g_{\Phi\Phi}(0)} \frac{(\ell /\ell^*)}{1 - (\ell/\ell^*)} 
\end{equation}
Using $g^2_{d\Phi}(0)/g_{\Phi\Phi}(0) = \gamma^2 g^2_{dd}(0)/[\gamma^2 g_{dd}(0)]\simeq \gamma^0  g_{dd}(0)$, the above
expression simplifies to:
\begin{equation}
g_{dd}(\ell) \simeq \frac{g_{dd}(0)}{1-(\ell/\ell^*)}.  
\end{equation}
which needs to be compared with the behavior of the renormalized coupling to the continuum after  setting $g_{\Phi\Phi}(0) \simeq \gamma^2 g_{dd}(0)$:
\begin{equation}
g_{\Phi\Phi}(\ell) \simeq \frac{\gamma^2 g_{dd}(0)}{1-(\ell/\ell^*)}.
\end{equation}
Note that both couplings diverge at $\ell^* = \log(D/T^{\Phi}_K)$
with $T^{\Phi}_{K}\simeq D e^{-1/2(\nu_0 \gamma^2 J_{dd})}\ll \Delta$ if $\gamma\ll 1$, which is consistent with what was  discussed above.  For instance, if we choose  $\gamma\approx 0.2$ (corresponding to $\Theta\approx 0.8$), then
\begin{equation}
\frac{g_{\Phi\Phi}(\ell)}{g_{dd}(\ell)} \simeq \gamma^2 \ll 1.
\end{equation}
 Thus, as the continuum states are integrated out, the impurity  remains 
most strongly coupled to the single site describing the dGSJ quasi-particle and 
therefore the single-site model remains an accurate description of the magnetic impurity on the 
proximitized thin film.

Let us close this section by pointing out some  potential problems with the scaling analysis described above. First of all, like the original poor man's scaling~\cite{Anderson_poorman}, the equations are obtained perturbatively. Therefore, the solutions to the scaling equations are valid provided the couplings remain small compared to unity. This is not a problem under the above assumptions because the scale where the couplings diverge $\ell^*$ is much smaller than the superconductor gap
and the scaling must be stopped at the scale of $\Delta$. 
%However, it could be a problem in other  regimes of the model not explored here. 
As we get closer to the gap scale, the superconducting correlations cannot be neglected, and taking them into account will modify the flows of the renormalized couplings. Nevertheless, we should interpret the above analysis as providing  information on the tendency of the high-energy continuum states to couple to the impurity in the presence of the coupling to the dGSJ state.  
In order to follow the renormalization of the coupling to the continuum from high to low energies, it would be desirable to carry out  calculations using the NRG and starting from a more microscopic description of the system, e.g.  using  model parameters obtained from first principle calculations.
Such calculation should provide a more quantitative assessment of the accuracy the single-site model  introduced in this work for proximitized films.

\section{Conclusions}

We have studied the YSR excitations in a thin metal film proximitized by a superconductor. This has been carried out by introducing a one-dimensional model of the metal film/superconductor substrate. We have discussed the spectrum of this model, which consists of subgap bound states known as de Gennes-Saint James (dGSJ) states.  We have shown that Andreeev-reflection at the metal/superconductor interface leads to a substantial spectral reorganization around and below the gap energy. Next, the spectrum of the system when a magnetic impurity is deposited on the metal film has been also described. Treating the impurity spin as a classical vector, we have found there is substantial overlap of the wavefunctions of the Yu-Shiba-Rusinov (YSR) and the dGSJ states. Motivated by these results,  a single-site model has been introduced.  This model replaces the complexity of the proximitized film with a single-site that represents the dGSJ quasi-particle excitation and is coupled to the impurity with an effecive change coupling.  The single-site model is exactly solvable and allows us to  go beyond the classical description of the impurity by treating its spin quantum mechanically. Finally, we have addressed the accuracy of the single-site model by phenomenologically re-introducing the coupling to the continuum of excitations of the proximitized film as a perturbation and using the poor man's scaling method: Under conditions suggested by the findings of  the classical approach, we have shown that  the  exchange coupling  with the site that describes the dGSJ quasi-particle excitation remains the dominant coupling under scaling.  Thus, the continuum of excitations of the proximitized film can be neglected in a first approximation, and the YSR states can be regarded as resulting from the exchange interaction  of the magnetic (quantum) impurity with the dGSJ quasi-particles.  

 The approach used here can be  generalized to treat  impurities with higher spin and account for single-ion as well as magnetic exchange anisotropies. Our results  provide  theoretical support for the model used to analyze the STS spectra reported in Ref.~\onlinecite{PhysRevLett.130.136004}. In addition, since the single-site  model introduced here is computationally cheaper than more sophisticated numerical methods like the numerical renormalization group (NRG)~\cite{zitko} or continuous-time Montecarlo~\cite{Odobesko_PhysRevB.102.174504}, it can be used
 to model more complex systems such as chains  or other nanostructures  of magnetic impurities on proximitized films, which would be otherwise rather intractable by those methods.  For this reason, we also believe it is worth revisiting the  system studied here using much more sophisticated numerical tools, in order to quantitatively  assess the limitations of the single-site model as introduced in this work. 

 \begin{acknowledgments}
We acknowledge financial support from Grants No. PID2019-107338RB-C61, No.  CEX2020-001038-M, No. PID2020-112811GB-I00, and  PID2020-114252GB-I00, funded by MCIN/AEI/ 10.13039/501100011033, from the Diputación Foral de Guipuzcoa, the ELKARTEK project BRTA QUANTUM (no. KK-2022/00041), and from the European Union (EU) through the Horizon 2020 FET-Open projects SPRING (No. 863098) and SUPERTED (No. 800923),  and the European Regional Development Fund (ERDF).
M.A.C. has been supported by Ikerbasque, Basque Foundation for Science, and MCIN Grant No. PID2020-120614GB-I00 (ENACT).
F.S.B. thanks Prof. Bj\"orn Trauzettel for his hospitality at W\"urzburg University,
and the A. v. Humboldt Foundation for financial support.
J.O. acknowledges the scholarship  PRE\_2022\_2\_00950  from the Basque Government.
The authors also thank Katerina Vaxevani and Stefano Trivini for their help measuring the experimental spectrum on Fig \ref{fig:1} (d) and several discussions.
\end{acknowledgments}

\onecolumngrid

\appendix

\section{Two-layer Green's Functions}\label{app:A}

As discussed in Ref.~\onlinecite{Feuchtwang}, the GFs of a composite system~\eqref{H0} can be obtained from the GFs of the constituent subsystems. We denote the GFs of each subsystem as $g_i(x,x')$, with $i=N,S$. Next, we impose the following boundary conditions:
\begin{align}
     \dfrac{d g_i(x,x')}{dx}\Big|_{x=0} &=0,\qquad  \dfrac{d g_i(x,x')}{dx'}\Big|_{x'=0} = 0,\\
    \dfrac{d g_N(x,x')}{dx}\Big|_{x=-a}&=0,\qquad   \dfrac{d g_N(x,x')}{dx'}\Big|_{x'=-a} =0,\\
    g_S(x\to +\infty,x')&=0,\qquad   g_S(x,x'\to +\infty) = 0,\\
     \lim_{\delta \to 0^+}\,\tau_3\dfrac{d g_i(x,x')}{dx}\Big|^{x=x'+\delta}_{x=x'-\delta}&=\dfrac{2m}{\hbar^2}, \qquad \lim_{\delta \to 0^+}\, \tau_3\dfrac{d g_i(x,x')}{dx'}\Big|^{x'=x+\delta}_{x'=x-\delta}=\dfrac{2m}{\hbar^2}.
\end{align}
We assume a zero derivative at the vacuum interface, except for the semi-infinite superconductor at $x,x'\rightarrow\infty$, for which the GFs are assumed to vanish. %The last condition is the jump condition. 

Using the above boundary conditions, and assuming continuity of the full GF and its derivative at the N/S interface, we obtain the following relations:
\begin{equation}\label{GT}
   G(x,x')= \left\{\begin{aligned}
    & g_S(x,x')\theta(x')\mp g_S(x,0)[g_S(0,0)+g_N(0,0)]^{-1}g_{S,N}(0,x')\text{ , for  } x>0, x'\gtrless 0 \\
    &g_N(x,x')\theta(-x')\mp g_N(x,0)[g_S(0,0)+g_N(0,0)]^{-1}g_{S,N}(0,x')\text{ , for  } x<0, x\gtrless 0 
    \end{aligned}\right.\; .
\end{equation}
The GFs for the isolated system are easy to calculate from Bogoliubov-de Gennes equation \cite{McMillan}. After that, one can get the dressed GF from \eqref{GT}, see [Arnold]. The GF shows a pole with the following energy distribution:
\begin{equation}\label{ESJdG}
    \dfrac{\omega}{\sqrt{\Delta^2-\omega^2}}\tan\left( \frac{2 m a}{\hbar^2 k_F}\omega\right)=1
\end{equation}
Note that by expanding the tangent around zero to the first order we arrive at the same solution as obtained from a semi-classical argument Eq.~\eqref{eq:dGsJ_0}.

\section{Single site Hamiltonian for a proximitized superconductor}\label{app:B}

By solving the Bogoliubov-de Gennes equations for the proximitized thin film, the
electron field operator at the position of the magnetic impurity,$\phi_{0\sigma}$,
can be written as follows:
\begin{equation}
    \psi_{0\sigma}=u_0\gamma_{\sigma}+\sigma v^*_0\gamma^{\dagger}_{-\sigma}+ \Phi_{o\sigma}.
    \label{eq:splitfield}
\end{equation}
We start by changing the basis on the unperturbed Hamiltonian \eqref{H0},
where first two terms described the dGSJ quasi-particle and $\Phi_{0\sigma}$ describes the 
modes in the continuum.   We now introduce a rotation for the operators creating the discrete state:
\begin{equation}
    \begin{pmatrix}
    d_{\sigma}\\
    d^{\dagger}_{-\sigma}
    \end{pmatrix}=\begin{pmatrix}
    \cos\theta& -\sin\theta \\
    \sin\theta & \cos\theta
    \end{pmatrix}\begin{pmatrix}
    \gamma_{0\sigma}\\
    \gamma^{\dagger}_{0-\sigma}
    \end{pmatrix}\; .
\end{equation}
This rotation leads to ~\eqref{eq:H0Sd}, where  $U=E_s\cos2\theta$ and $\Delta_s=E_s\sin2\theta$. Furthermore, requiring that 
\begin{equation}  
\psi_{\uparrow}=\sqrt{Z}d_{\uparrow}+ +\Phi_{0\uparrow} =u_0\gamma_{\sigma}+\sigma v^*_0\gamma^{\dagger}_{-\sigma}+\Phi_{0\uparrow}\;.
\end{equation}
Hence, $\tan\theta=-v_0/u_0$ and $Z=u_0^2+v_0^2$, where $U=E_s (u_0^2-v_0^2)$ and $\Delta_s=2E_s u_0v_0$.

\section{Calculation of the scaling equations}\label{app:C}

 In order to perturbatively obtain the scaling equations of the model introduced 
 in Sec.~\ref{sec:PoorMan}, we consider an expansion of the partition function  of the system, i.e.
 \begin{equation}
 Z(D) = Z_0(D) \big\langle \mathcal{T} \exp\left[ -\int^{\beta}_0 H^{dc}_{J}(\tau) \right] \big\rangle_0,
 \end{equation}
 in powers of the couplings $J_{dd},J_{d\Phi}$ and $J_{\Phi\Phi}$. 
 In the above expression for $Z(D)$ $Z_0(D) = \mathrm{Tr}\: e^{\beta H_0}$ is the 
 partition function of the system without
 magnetic impurity at inverse absolute temperature $\beta = (k_B T)^{-1}$;
 $\langle \ldots \rangle_0$ is the expectation value over the non-interacting grand canonical ensemble defined by $H_0$. The operator $H_J(\tau) = e^{H_0 \tau} H_J e^{-H_0\tau}$,
 where $H_J$ is given in Eq.~\eqref{eq:phenolham}, describes the magnetic exchange with the impurity  in the interaction representation and $\mathcal{T}$ is the imaginary time-ordering symbol. We have also introduced a parameter, $D~E_F$, which is the bandwidth of the composite thin film and superconductor system.
 
  Following Anderson~\cite{Anderson_poorman}, we shall use  perturbation theory to obtain a map
  onto a system with smaller bandwidth $D^{\prime}  = D -\delta D < D$. Associated with the bandwidths $D$ and
  $D^{\prime}$,   there are also the following characteristic (imaginary) time scale (in units where $\hbar = 1$) $\tau_c = D^{-1}$ and $\tau^{\prime}_c = (D^{\prime})^{-1} > \tau_c$. The lowest order terms of the perturbation series for the system with bandwidth $D$ read:
\begin{equation}
Z(D) = Z_0(D) \left\{ 1  - \int d\tau\, \langle H^{dc}_J(\tau) \rangle_0  + \frac{1}{2!} \int\limits_{|\tau-\tau^{\prime}| > \tau_c= D^{-1}} 
d\tau d\tau^{\prime}\, \langle T\left[ H^{dc}_J(\tau) H^{dc}_J(\tau^{\prime}) \right]\rangle_0 + \cdots 
\right\},\label{eq:pert}
\end{equation}
 where we have made explicit the constraints on $\tau$ imposed by the finite bandwidth of the continuum of states described by $\Phi_{\sigma}$ and $\Phi^{\dag}_{\sigma}$.

  Next, let us integrate out the high energy degrees of freedom contained $\Phi_{0\sigma}$ and $\Phi^{\dag}_{0\sigma}$ (recall that $d_{\sigma}, d^{\dag}_{\sigma}$ describe a low-energy subgap state and it cannot be integrated out). Such degrees of freedom involve excitations with energies $\sim D$ above the ground   state and therefore determine the short imaginary time behavior of the Green's functions for $\Phi_{\sigma}$. Note that, since at excitation energies $\sim D$
  Bogoliubov quasi-particles either behave as electrons or holes (in other words, either 
  $u\to 0$ or $v\to 0$), the anomalous  GFs involving the operator $\Phi_0$, ie. 
  $\langle \mathcal{T}\left[ \Phi_{0\uparrow}(\tau) \Phi_{0\downarrow} (\tau^{\prime}) \right] \rangle_0 $, etc, vanish  for $|\tau-\tau^{\prime}| \simeq \tau^{-1}_c$. 
  Thus, in the above perturbation series,  for $|\tau-\tau^{\prime}|\sim \tau_c$, we need to consider only   normal correlations, which take the familiar Fermi liquid form:
  \begin{equation}
  \langle \mathcal{T}\left[ \Phi_{0\sigma}(\tau) \Phi^{\dag}_{0\sigma^{\prime}} (\tau^{\prime}) \right] 
  \rangle_0 \simeq  \frac{ \nu_0\delta_{\sigma\sigma^{\prime}}}{(\tau-\tau^{\prime})}
  \end{equation}
  for $|\tau^{\prime}-\tau| \simeq \tau^{-1}_c$, where $\nu_0$ is the (mean) density of states of the normal state. Thus, the first non-constant contribution to the scaling of the
  couplings  stems from the second order term. We first split the integrals over $\tau,\tau^{\prime}$ according to:
  \begin{equation}
  \int\limits_{|\tau-\tau^{\prime}| > \tau_c= D^{-1}} 
d\tau d\tau^{\prime}\,  \ldots  = \int\limits_{|\tau-\tau^{\prime}| > \tau^{\prime}_c= (D^{\prime})^{-1}} 
d\tau d\tau^{\prime}\,  \ldots  + \int\limits_{\tau^{\prime}_c = (D^{\prime})^{-1} > |\tau-\tau^{\prime}| > \tau_c = D^{-1}} 
d\tau d\tau^{\prime}\,  \ldots
  \end{equation}
 and consider the terms in the second term for which $\tau^{\prime}_c > |\tau-\tau^{\prime}| > \tau_c$. Expanding the second order term in powers, corrections to the
 couplings contained in the first order term are generated at $O(J^2_{d\Phi})$, $O(J_{d\Phi} J_{\Phi\Phi})$ and $O(J^{2}_{\Phi\Phi})$.  We explicity evaluate below the $O(J^2_{d\Phi})$ term. The calculations for the remaining terms are similar  and  not reproduced here. Einstein's convention of repeated index summation is used throughout:
 \begin{align}
 O(J^2_{d\Phi}) &=  \frac{J^2_{d\Phi}(D)}{2!} \int\limits_{\tau^{\prime}_c > |\tau-\tau^{\prime}| > \tau_c} 
d\tau d\tau^{\prime}\, \left\{ \langle \mathcal{T} \left[ S^a(\tau) S^b(\tau^{\prime})\right] \rangle_0 
  \left( s^a_{\sigma\sigma^{\prime}} s^b_{\lambda\lambda^{\prime}} \right) \langle \mathcal{T}\left[  d^{\dag}_{\sigma}(\tau)  \Phi_{0\sigma^{\prime}}(\tau) \Phi^{\dag}_{0\lambda}(\tau^{\prime}) d_{\lambda^{\prime}}(\tau^{\prime}) \right] \rangle_0 \right. \notag \\
   & \qquad \qquad \qquad \qquad  +  \left.   \langle \mathcal{T} \left[ S^a(\tau) S^b(\tau^{\prime})\right] \rangle_0 
  \left( s^a_{\sigma\sigma^{\prime}} s^b_{\lambda\lambda^{\prime}} \right) 
  \langle \mathcal{T}\left[    \Phi^{\dag}_{0\sigma}(\tau) d_{\sigma^{\prime}}(\tau)  d^{\dag}_{\lambda}(\tau^{\prime})  \Phi_{0\lambda^{\prime}}(\tau^{\prime})\right] \rangle_0 \right\}  \\
  &=  
   \frac{J^2_{d\Phi}(D)\nu_0}{2!} \int\limits_{\tau^{\prime}_c > |\tau-\tau^{\prime}| > \tau_c} 
d\tau d\tau^{\prime}\, \left\{ \langle \mathcal{T} \left[ S^a(\tau) S^b(\tau^{\prime})\right] \rangle_0 
   \frac{ \left(s^a_{\sigma\lambda} s^b_{\lambda\lambda^{\prime}} \right)}{(\tau-\tau^{\prime})} \langle \mathcal{T}\left[  d^{\dag}_{\sigma}(\tau)  d_{\lambda^{\prime}}(\tau^{\prime}) \right] \rangle_0 \right. \notag \\
   & \qquad \qquad \qquad \qquad  + \left.   \langle \mathcal{T} \left[ S^a(\tau) S^b(\tau^{\prime})\right] \rangle_0 
   \frac{ \left(s^b_{\lambda\sigma} s^a_{\sigma\sigma^{\prime}}  \right) }{(\tau-\tau^{\prime})}
  \langle \mathcal{T}\left[     d_{\sigma^{\prime}}(\tau)  d^{\dag}_{\lambda}(\tau^{\prime}) \right] \rangle_0 \right\}  \\
  &=  - \frac{J^2_{d\Phi}(D)\nu_0}{4} \int\limits_{\tau^{\prime}_c > |\tau-\tau^{\prime}| > \tau_c} d\tau d\tau^{\prime} \,   \frac{i\epsilon^{abc} \left[s^a, s^b\right]_{\sigma\sigma^{\prime}}  }{|\tau-\tau^{\prime}|} \langle T\left[ S^c(\tau)\right] \rangle_0 
  \langle \mathcal{T}\left[ d^{\dag}_{\sigma}(\tau)  d_{\sigma^{\prime}}(\tau^{\prime}) \right] \rangle_0 \\
  &=  - \frac{J^2_{d\Phi}(D)\nu_0}{2} \int\limits_{\tau^{\prime}_c > |\tau-\tau^{\prime}| > \tau_c} d\tau d\tau^{\prime} \,   
  \frac{\left(  \epsilon^{abc} \epsilon^{abf} s^f_{\sigma\sigma^{\prime}} \right) }{|\tau-\tau^{\prime}|} 
   \langle T\left[ S^c(\tau)\right] \rangle_0 
  \langle \mathcal{T}\left[ d^{\dag}_{\sigma}(\tau)  d_{\sigma^{\prime}}(\tau^{\prime}) \right] \rangle_0\\
 &=- \frac{J^2_{d\Phi}(D)\nu_0}{2} \int\limits_{\tau^{\prime}_c > |\tau-\tau^{\prime}| > \tau_c} d\tau d\tau^{\prime} \,   
  \frac{1}{|\tau-\tau^{\prime}|} \, 
  \langle \mathcal{T}\left[ d^{\dag}_{\sigma}(\tau) S^c(\tau) s^{c}_{\sigma\sigma^{\prime}} d_{\sigma^{\prime}}(\tau^{\prime}) \right] \rangle_0
 \end{align}
In the above derivation we have used the following results: $\epsilon^{abc}\epsilon^{abd} = 2\delta_{cd}$ and
\begin{align}
 \mathcal{T} \left[ S^a(\tau) S^b(\tau^{\prime})\right] &= 
\theta(\tau-\tau^{\prime})  S^a S^b + \theta(\tau^{\prime}-\tau)
 S^b S^a \\
&= \frac{1}{2} (S^a S^{b} - S^{b}S^{a}) \left[\theta(\tau-\tau^{\prime}) - \theta(\tau^{\prime}-\tau) \right]   + \frac{1}{2} (S^a S^{b} + S^{b}S^{a}) \\
&= \frac{i}{2} \epsilon^{abc}S^c \mathrm{sgn}(\tau-\tau^{\prime}) + \left\{ S^{a},S^{b} \right\}  
\end{align}
because $S^a (\tau) = e^{H_0\tau} S^a e^{-H_0 \tau} = S^a$. As noted above, 
 the operators describing the dGSJ quasi-particle have time dynamics varying on the scale of $\Delta^{-1}\ll \tau^{\prime}_c$, which is very slow compared to the fast degrees of freedom  being integrated out from $\Phi_{0c}$ and $\Phi^{\dag}_c$. Introducing $\tau_{-} = \tau -\tau^{\prime}$ and $\tau_{+} = (\tau+\tau^{\prime})/2$. Thus,  the term proportional to
$\{S^a,S^b\}$ drops because it is multiplied by $\tau_{-}^{-1}$ rather
than $|\tau_{-}|^{-1}$ and the integral over $\tau_{-}$ of former 
vanishes to leading order. Thus, to leading order in $\tau_{-}$,  we are left with
\begin{align}
O(J^{2}_{d\Phi}) &= -J^2_{d\Phi}(D) \nu_0\int d\tau_{+} \, \langle \mathcal{T} \left[ d^{\dag}_{\sigma}(\tau_{+}) \vec{S}(\tau_{+})\cdot \vec{s}_{\sigma\sigma^{\prime}}(\tau_{+}) d_{\sigma}(\tau_{+}) \right] \rangle_0  \, \int\limits_{\tau^{\prime}_c > |\tau_{-}| > \tau_c} \frac{d\tau_{-}}{|\tau_{-}|}
\\
&=  -2  \nu_0  \frac{\delta D}{D}  J^2_{d\Phi}(D)  \int d\tau \, \langle \mathcal{T} \left[ d^{\dag}_{\sigma}(\tau) \vec{S}(\tau)\cdot \vec{s}_{\sigma\sigma^{\prime}}(\tau) d_{\sigma}(\tau) \right] \rangle_0. \label{eq:res}
\end{align}
In the last expression, we have evaluated the integral over  $\tau_{-}$ using
\begin{equation}
\int\limits_{\tau^{\prime}_c > |\tau_{-}| > \tau_c} \frac{d\tau_{-}}{|\tau_{-}|} = 2 \log\left(\frac{\tau^{\prime}_c}{\tau_{c}} \right) = 2 \log\left(\frac{D}{D^{\prime}} \right) = 
-2 \log \left( \frac{D-\delta D}{D} \right) \simeq\frac{2\delta D}{D},
\end{equation}
and replaced $\tau_{+}\to \tau$. Notice that the resulting expression in Eq.~\eqref{eq:res} takes the same form as the contribution $\propto J_{dd}$ in the first order term of \eqref{eq:pert}. This leads to the following  recursion relation:
\begin{align}
J_{dd}(D-\delta D) &= 
J_{dd}(D)  + 2 \nu_0  J^2_{d\Phi}(D) \frac{\delta D}{D}\\
\end{align}
Assuming the couplings are continuous functions of the cut-off $D$, the recursion relation
becomes a differential equation:
\begin{equation}
D \frac{dJ_{dd}(D)}{d D} = - 2 \nu_0 J^2_{d\Phi},
\end{equation}
which implies that $J_{cc}$  increases with decreasing  bandwdith $D$. 

Similarly, we can tackle the terms  at $O(J_{d\Phi}J_{\Phi\Phi})$ and $O(J_{\Phi\Phi})$ (note the
latter one is the  only one present in the standard poor man's scaling treatment of the Kondo
model). From those terms, the following differential equations are obtained:
\begin{align}
D \frac{dJ_{d\Phi}(D)}{d D} &= - 2 \nu_0 J_{d\Phi} J_{\Phi\Phi},\\
D \frac{dJ_{dd}(D)}{d D} &= - 2 \nu_0 J^2_{\Phi\Phi}.
\end{align}
It is convenient to introduce a new scaling variable defined by the differential
equation:
\begin{equation}
\frac{dD}{D} = d\ell \Rightarrow D(\ell) = D_0 e^{-\ell}.
\end{equation}
Thus, as $\ell\to +\infty$ $D(\ell)\to 0$. Furthermore,
if we define the dimensionless couplings $g_{dd} = 2\nu_0 J_{dd}$,
$g_{d\Phi} = 2\nu_0 J_{d\Phi}$, and $g_{\Phi\Phi} =2\nu_0 J_{\Phi \Phi}$,
we finally arrive at the scaling equations~\eqref{eq:rg1} to \eqref{eq:rg3}
discussed in Sec.~\ref{sec:PoorMan}.

\twocolumngrid

%\bibliographystyle{apsrev4-1} 

%\bibliography{references}


\end{document}